# Status of the OPERA neutrino experiment

H. Pessard
on behalf of the OPERA Collaboration

**LAPP - Université de Savoie - CNRS/ IN2P3**
BP. 110, F-74941 Annecy-le-Vieux Cedex, France
*E-mail: henri.pessard@lapp.in2p3.fr*

The OPERA long-baseline neutrino oscillation experiment is located in the underground Gran Sasso laboratory in Italy. OPERA has been designed to give the first direct proof of $\nu_\tau$ appearance, looking at the CNGS $\nu_\mu$ beam 730 km away from its source at CERN. The apparatus consists of a large set of emulsion-lead targets combined with electronic detectors. First runs in 2007 and 2008 helped checking that detector and related emulsion facilities are fully operational and led to successful first analysis of collected data. This paper, after a short description of the OPERA setup and methods, gives an updated status report on data reconstruction and analysis applied to available samples of neutrino events

## 1. The CNGS program and OPERA

Following the discovery of oscillations with atmospheric neutrinos by Super-Kamiokande in 1998, accelerator neutrino projects developed in Japan and the USA to measure the $\nu_\mu$ disappearance at long distance. The $\nu_\mu$ disappearance signal is now confirmed by K2K and MINOS. In Europe, long baseline projects focused on the $\nu_\tau$ appearance in a $\nu_\mu$ beam. They led to the construction of the CNGS (CERN Neutrinos to Gran Sasso) beam at CERN, whose main physics objective is to prove explicitly the $\nu_\mu - \nu_\tau$ nature of the atmospheric oscillation.

To achieve this purpose, the OPERA experiment, using targets made with lead and nuclear emulsions, was developed by an international collaboration of 12 countries and ~200 physicists to operate in the Gran Sasso underground laboratory, 730 km away from CERN. Construction started in 2003. The experimental setup was completed in 2008 and is now in operation.

The CNGS beam is a dedicated wide band neutrino beam, with a mean energy of 17 GeV optimized for $\nu_\tau$ production and detection. New SPS beam extraction, target and focusing horns, km long decay tunnel have been constructed from 2000 to 2006. With a foreseen intensity of $4.5\ 10^{19}$ 400 GeV *protons on target* (p.o.t.) per year (assuming 200 days of operation), the number of neutrino interactions expected to occur in 1.25 kiloton of OPERA targets in 5 years is about 24000. The number of $\nu_\tau$ CC interactions is then about 120 for $\Delta m^2 = 2.5\ 10^{-3}$ $eV^2$, leading to the observation of about 10 $\nu_\tau$ CC events with less than one background event.

## 2. OPERA detector principles

The OPERA experiment is based on the direct observation of the $\tau$ decay topology in nuclear emulsion films, the only technique providing micron level spatial resolution. Charged tracks produce ~32 grains /100 μm of emulsion, allowing reconstruction accuracies of order 1 μm in position and 1 mrad in angle. The detection of $\nu_\tau$ CC events by the $\tau$ decay kink and/or the impact parameter of the daughter in 1-prong $\tau$ decay modes (84% total branching ratio), and by the displaced $\tau$ vertex in the 3-prong $\tau$ decay mode (15%) is based on emulsion films for tracking, associated with high density plates as targets. This technique, used for the $\nu_\tau$ discovery by DONUT in 2000, was brought to an industrial scale in OPERA (100 000 $m^2$ of films).

The necessity of keeping film alignment within micron precision while achieving a target mass of more than a kiloton led to design highly modular targets. The basic target unit is a *brick* of 10.2 x 12.8 x 7.9 cm made of 56 lead plates and 57 emulsion films. The emulsion films are made of a 205 μm plastic base with two emulsion layers of 44 μm, they alternate with 1 mm thick lead plates. Each brick (Fig. 1, left) has a removable pair of *Changeable Sheets* (CS) films on its downstream face. There are altogether 150 000 bricks for a total mass of 1.25 kiloton.

The OPERA detector [1] is composed of two identical parts, each made of a target section and a muon spectrometer (Fig. 1, right), with an upstream large size RPC veto detector. In each target section, steel trays able to hold brick rows in vertical *walls* alternate with 31 planes of electronic detectors, the *Target Trackers* (TT) made of crossed 7 m long scintillator strips read at both ends via WLS fibers by 64-channel PMT. TT allow triggering on neutrino interactions and localizing the bricks in which they occured. Each muon spectrometer is composed of a large dipolar iron magnet instrumented with 22 RPC planes and complemented by 6 sections of drift tubes for precision tracking through the magnetic field. These detectors are used to identify muons emerging from the interactions and measure their momentum and charge, an important feature to distinguish background events containing $\nu_\mu$-produced charm particles.

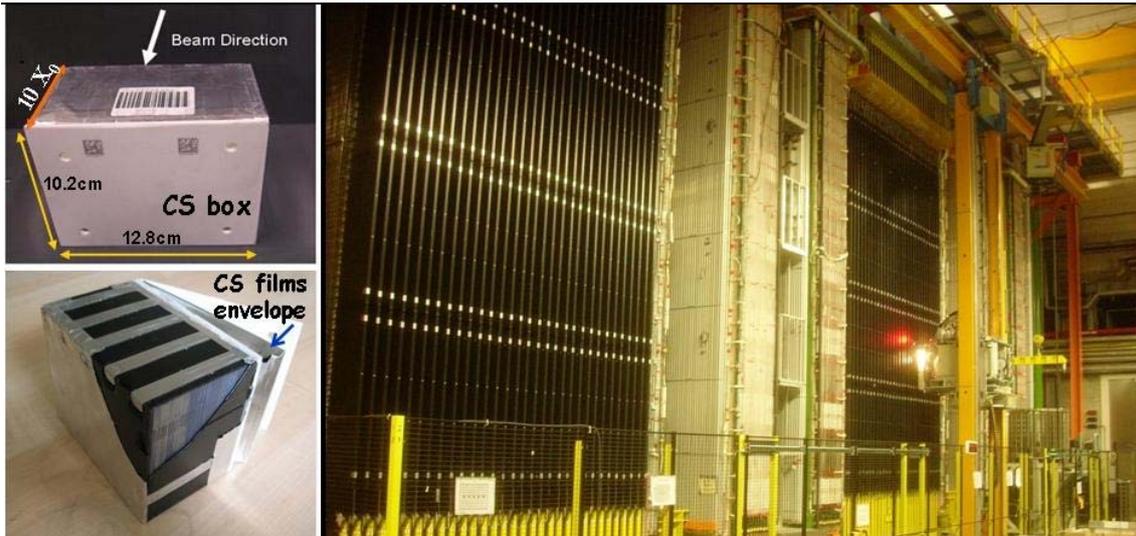

**Figure 1.** On the left are shown two pictures of OPERA target bricks, a complete brick (top) and an exploded view of a brick (bottom) showing its structure and the CS films packaging. On the right, a picture of the OPERA detector shows the two target sections (black covers) and their spectrometers. One of the brick manipulators is seen operating along the second target block.

### 3. The OPERA event flow and analysis

The OPERA data acquisition is synchronized with the CNGS beam by a double GPS system with an accuracy of 100 ns, leaving negligible cosmic background among on-time events. After event track reconstruction in the TT and spectrometer detectors, interactions occurring in target areas are selected and the probability to contain the neutrino interaction vertex is computed for surrounding bricks. Highest probability bricks designated by the brick finding algorithm are extracted continuously by the *Brick Manipulator System* (BMS) and exposed to frontal X-rays to create alignment references between CS and brick films.

CS films are detached and analysed to validate the brick selection and provide predictions for the event localization. If a TT-predicted track is found in the CS, the brick is exposed to lateral X-rays, then to high energy cosmic rays in the surface laboratory, to generate alignment tracks. The brick is then unpacked, its emulsion films are developed and sent to the scanning laboratories of the Collaboration for vertex location and further data collection in the emulsions relevant for decay searches. If no matching track is found in the CS, the brick is returned to the detector with new CS and the next brick in the probability list is extracted.

Automated emulsion scanning of CS and brick films is shared between Europe and Japan. CS scanning is centralized in two laboratories, at Gran Sasso (LNGS) and at Nagoya University. Scanning is done with computer driven microscopes. CMOS cameras and commercial hardware are used with software algorithms in the microscope system common to the nine European scanning labs. Fewer but faster (up to 75 $cm^2$/h compared to 20 $cm^2$/h) systems using high-speed CCD cameras with custom mechanics and hard-coded algorithms are exploited in Japan.

Tracks found in the CS are searched for in the most downstream film of the brick and followed back in the preceding films. Track stopping indicates a possible vertex. To confirm and study a vertex, an area of about 1 $cm^2$ is measured in 5 films upstream and 5 films downstream. All tracks found in this volume are input to a vertex reconstruction program tuned to find decay topologies. To complete the kinematical analysis of the event, a momentum measurement algorithm using Multiple Coulomb Scattering in the lead plates is applied for low energy tracks. In addition, electromagnetic showers developing in the plates can be measured, their direction and energy reconstructed, and electrons identified.

## 4. OPERA running and first results

The OPERA electronic detector construction was completed in spring 2007. After a research period devoted to brick packaging and lead-emulsion compatibility studies, 150 000 bricks have been assembled from March 2007 to June 2008 by an automated *Brick Assembly Machine* located in the underground laboratory. Produced at an average rate of 700 per day, bricks have been continuously inserted into the detector by the BMS during the same period.

After a short commissioning run in 2006 without bricks, the CNGS operation started in September 2007 with 40% of the OPERA bricks. Due to radiation shielding problems in the CNGS target zone, this run lasted also just a few days, 38 interactions were recorded in the bricks for 0.08 $10^{19}$ protons delivered on target. The first runs allowed checking the beam time synchronization [2] and the reconstruction of CNGS events with the electronic detectors.

A first long OPERA physics run took place from June to November 2008, although poor performance in the first three months limited the average efficiency of the CNGS to about 60%. OPERA recorded 10100 events and among them ~1700 interactions in the targets for 1.78 $10^{19}$ p.o.t. delivered. All subdetectors were operational and the live time of the data acquisition system was ~99%. Brick extraction, handling of bricks and CS, CS development and scanning, brick dismounting, film development and shipping, film scanning, all large engineering tasks, were validated at nominal speed during the 2008 run.

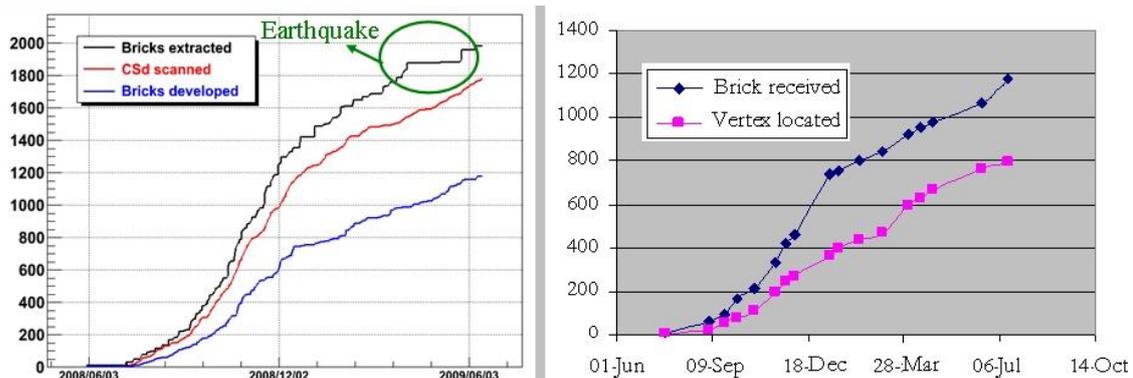

**Figure 2.** Statistics of the brick processing activities at the time of the EPS-HEP 2009 conference are shown for the 2008 sample. Fluxes at LNGS and CS scanning centers (left plot) and at brick scanning laboratories (right plot) can be seen.

Following improvements in reconstruction and analysis procedures [3], successive steps of the data collection from the bricks developed in a progressive manner, as illustrated by the brick processing statistics (Figure 2). At the time of the conference, 797 interactions vertices had been located (119 '0-μ' and 678 '1-μ' interactions), allowing to start decay searches.

The CS principle worked as expected to perform efficient track matching between TT and brick emulsion films. Brick finding rate after the $2^{nd}$ brick is presently ~80%. Work in progress to improve brick prediction algorithms will benefit from the publication on database of larger statistics. Investigation of background and efficiencies from the data will also become possible.

The decay search is ongoing in the 2008 sample and a number of interesting events have already been found, namely charm decay candidates. Charm events, with topologies similar to τ events, are crucial to cross-check τ identification efficiency. Figure 3 shows a charm candidate as seen in the emulsions and illustrates important progress achieved in the last months to identify short distance decays, by recovering tracks found in only one of the two emulsion layers of the film. The method improves also significantly the track impact parameter distribution in vertices, as shown in Figure 3, bottom right.

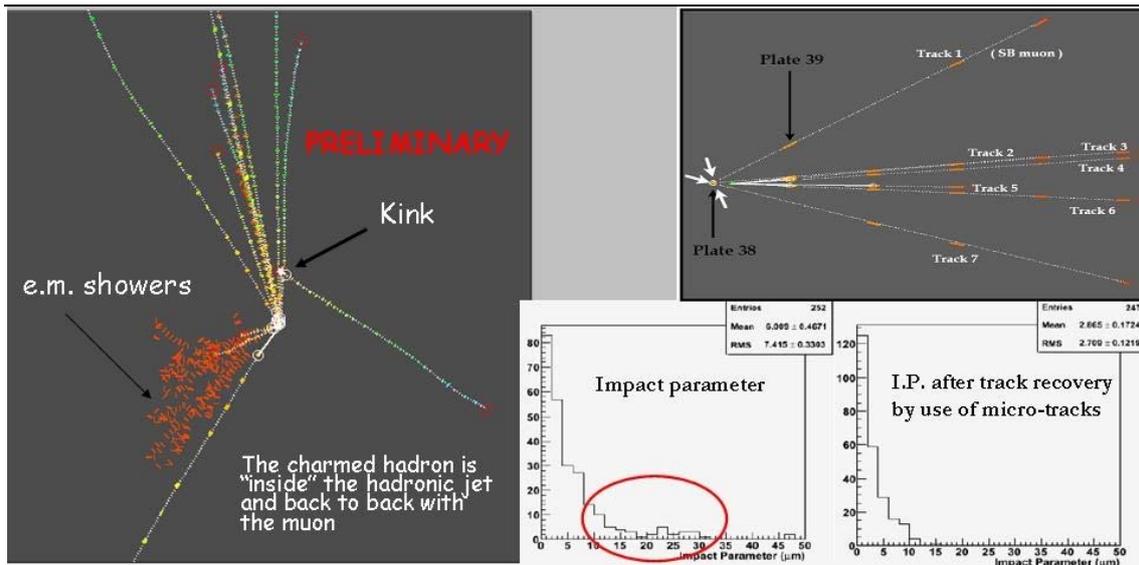

**Figure 3.** An example of charm decay candidate with a single prong kink (left). A neutral charm decay candidate (top right) found after recovery of tracks found in only one of the two emulsion layers of the film, using single micro-tracks. Bottom right plots show the improvement of the distribution of impact parameters after this method was applied.

In 2009, a few weeks before the foreseen start of the CNGS, a strong earthquake affected the area of Gran Sasso hitting severely the city and villages of L'Aquila. The LNGS experiments suffered only minor damages. The most important impact was on residents. Local activities, run preparation and processing of 2008 candidates, were stopped for more than a month. The CNGS run could nevertheless be started with only two weeks delay on June 1$^{st}$. Since then, the beam has been delivered with rather good efficiency, letting expect the order of 3.5 $10^{19}$ p.o.t. for the whole period. Processing of the 2009 sample has immediately started, in parallel with the end of the 2008 analysis.

**5. Conclusions**

After two short runs in 2006 and 2007, OPERA started significant data taking in 2008 when about 1700 neutrinos interacted in the target bricks. The Collaboration is finishing the extraction of data from the emulsions and their analysis, locating the final sample of vertices for $\nu_\tau$ candidate search. Charm induced events study will be an important tool to improve control of efficiencies and background. A new physics run has started in June 2009, 3.5 $10^{19}$ protons on target could be delivered in about 160 days and ~3600 events collected in the OPERA bricks. About 2 $\tau$'s will then be expected in the total sample 2008-2009, opening the way to the direct observation of $\nu_\mu$–$\nu_\tau$ oscillations. To reach quickly its physics goal, OPERA relies on the CERN support to have the CNGS running at nominal intensity in the next years.